\documentclass{llncs}
\usepackage{times,graphicx,hyperref}
\usepackage{float}
\usepackage[section]{placeins}

\begin{document}
\title{POS Tagging and its Applications for Mathematics}
\subtitle{Text analysis in mathematics}
\author{Ulf Sch\"{o}neberg\inst{} \and Wolfram Sperber\inst{}}
\authorrunning{Ulf Sch\"{o}neberg, Wolfram Sperber}
\tocauthor{Ulf Sch\"{o}neberg and Wolfram Sperber}
\institute{FIZ Karlsruhe/Zentralblatt MATH, Franklinstr. 11, 10587 Berlin, Germany}
\maketitle

\begin{abstract}

Content analysis of scientific publications is a nontrivial task, but a useful and important one for scientific information services.  In the Gutenberg era it was a domain of human experts; in the digital age many machine-based methods, e.g., graph analysis tools and machine-learning techniques, have been developed for it.  Natural Language Processing (NLP) is a powerful machine-learning approach to semiautomatic speech and language processing, which is also applicable to mathematics. The well established methods of NLP  have to be adjusted for the special needs of mathematics, in particular for handling mathematical formulae. We demonstrate a mathematics-aware part of speech tagger and give a short overview about our adaptation of NLP methods for mathematical publications. We show the use of the  tools developed for key phrase extraction and classification in the database zbMATH.     

%keywords{Part of Speech (POS) Tagging, natural language processing, keyword extraction, classification}
%\keywords{POS tagging, Part of Speech Tagging, natural language processing}
\end{abstract}

\section{Methods and Tools}

\subsection{Part of Speech Tagging and Noun Phrases}
We describe our approach for Part Of Speech (POS) tagging and Noun Phrases (NPs) extraction of mathematical documents as basic tool for key phrase identification and classification of mathematical publications.
NLP methods arising from the field of computer linguistics constitute a statistics- and rule-based machine-learning approach to the processing of speech and language. Natural language analysis and understanding is a central aim of NLP. In western languages, Noun Phrases (NPs) are the most significant parts of sentences.
Extraction of the NPs and finding rules for which of them are relevant are the key aspects of automatic key phrase extraction in documents.

An important part of capturing NPs from a text is POS tagging. It presupposes the availability of information about the tokens in a sentence, especially the linguistic types of the tokens. Almost all state-of-the-art POS taggers rely on dictionaries. Some NLP tools are provided as Open Source software. In our project, the Stanford POS tagger~\cite{StanfordTag} is used. We extended the dictionaries of the POS tagger with large amounts of mathematical text data. We put a lot of work into these dictionaries as we already mentioned~\cite{delivermath} at CICM 2013. Especially, they contain names of mathematicians, acronyms and special terms that only exist in the domain of mathematics.

The Stanford tagger uses the Penn Treebank POS scheme~\cite{POS}, a classification scheme of linguistic types with 45 tags for tokens and punctuation symbols. This scheme has a relevant drawback for mathematical texts: there is no special tag for mathematical symbols or
mathematical formulae (in the following we subsume both to mathematical formulae). We did not change that. Formulae are handled by an auxiliary construct. This simple and straightforward method allows a slim and easy maintenance of the POS tagging software. 

In our approach, mathematical formulae (which are available as TeX code) are transformed to unique but random character sequences.

POS tagging has two main problems: new words and the ambiguity of POS tags (many tokens of the corpus can belong to more than one word class) are addressed by determining a suitable POS tag of a token using contextual statistical models and the Viterbi algorithm, a dynamic programming technique. The Viterbi algorithm uses information about the surrounding tokens to predict the probable POS tag of the ambiguous or unknown token, e.g., a formula is mainly tagged as an adjective or a noun. 

We illustrate that sequence of substitutions in the following example. It starts with the original LaTeX sentence and ends with the tagged sequence.
\small
The original sentence:

\begin{verbatim}
The classical Peano theorem states that in finite dimensional
spaces the Cauchy problem $x'(t)=f(t,x(t))$, $x(t\sb 0)=x\sb 0$,
has a solution provided $f$ is continuous.
\end{verbatim}

The TeX formulae are translated into unique, but randomly generated, character sequences:

\begin{verbatim}
The classical Peano theorem states that in finite dimensional
spaces the Cauchy problem formula-kqnompjyomsqomppsk,
formula-kqomolugwpjqolugwk, has a solution provided formula-kyk
is continuous.
\end{verbatim}

This sentence is fed into the POS tagger.
The Stanford tagger assigns an appropriate tag to ervey token.

\begin{verbatim}
The_DT Classical_JJ Peano_NNP theorem_NN states_VBZ that_IN 
in_IN finite_JJ dimensional_JJ spaces_NNS the_DT Cauchy_NNP
problem_NN formula-kqnompjyomsqomppsk_NN,
formula-kqomolugwpjqolugwk_NN ,_, has_VBZ a_DT solution_NN
provided_VBN formula-kyk_NN is_VBZ continuous_JJ ._.
\end{verbatim}
The tagged text is transformed back to its TeX representation without touching the tags:
\begin{verbatim}
The_DT Classical_JJ Peano_NNP theorem_NN states_VBZ that_IN
in_IN finite_JJ dimensional_JJ spaces_NNS the_DT Cauchy_NNP
problem_NN $x'(t)=f(t,x(t))$_NN,
$x(t\sb 0)=x\sb 0$_NN ,_, has_VBZ a_DT solution_NN
provided_VBN $k$_NN is_VBZ continuous_JJ ._.
\end{verbatim}
\normalsize

NP extraction is done by chunking with regular expressions for special patterns of POS tags. A very basic form of such a regular expression is: 

\small
\begin{verbatim}<DT>?<JJ>*<NN>\end{verbatim}
\normalsize 
This rule means that an NP chunk should be formed whenever the chunker finds an optional determiner (DT) followed by any number of adjectives (JJ) and then a noun (NN). However, it is easy to find many more complicated examples which this rule will not cover. The actual set of expressions we used is much more complex. Especially, NPs can be combinations of tokens and formulae, e.g., 'the Cauchy problem $x'(t)=f(t,x(t)),$ $x(t\sb 0)=x\sb 0$'.
After identifying the NPs in a document, we get a collection of NPs in the first step. 

If you would like to experiment with our solution we provide a web-based demo at
\url{http://www.zentralblatt-math.org/mathsearch/rs/postagger}

\subsection{Noun Phrase and Key Phrase Extraction}

A key phrase in our context, information retrieval in the mathematical literature, is a {\sl phrase that captures the essence of the topic of a document}~\cite{Wikipedia}. 

Mathematical publications, especially journal articles, have a more or less standardized metadata-structure covering 
important bibliographic data: authors, title, abstract, keywords (key phrases) and sometimes also a classification 
corresponding to the Mathematical Subject Classification (MSC).
Typically, key phrases are short phrases characterizing
\begin{itemize}
\item {embedding a publication in its general mathematical context as {\sl Diophantine equations} or {\sl optimal control}}
\item {special objects, methods, and results of a publication as {\sl bipartite complex networks}, {\sl k-centroids clustering}}
\end{itemize}

Often key phrases are descended form the title, the abstract or review or the fulltext, but this is not mandatory. key phrases of a document must not be part of the document.

NPs are natural candidates for key phrases. Key phrase identification via noun phrases is an usual technique.~\cite{NPKey}

\subsection{Classification with NPs}

Classification is also an important task within NLP. The normal approach which uses the full text of a document and favours stemming and Term Frequency/Inverse Document Frequency (TF/IDF) to get rid of redundant words, but we chose a different approach. We use the  extracted noun phrases from the texts, and than apply  text classification methods. We tested several machine learning techniques.  The best results were provided by a Support Vector Machine (SVM). The SVM we used is John Platt's sequential minimal optimization algorithm for support vector classifiers, the kernel is a polynomial kernel, the training data was every item from the database zbMATH from the beginning to the end of 2013. In particular, we used the Sequential Minimal Optimisation (SMO) technique from WEKA. 

\begin{figure}[h!]
\centering
\includegraphics[width=0.3\textwidth]{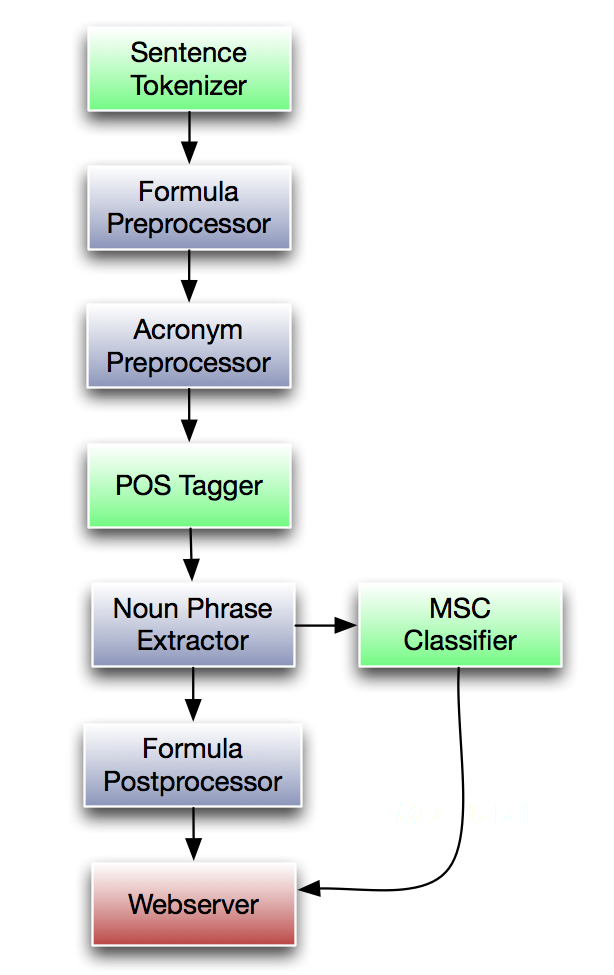}
\caption{The tagger with it's related processes}
\label{postagbuildblock}
\end{figure}

\subsection{The Big Picture}

A few words about the {\bf Fig. 1} block diagram: We start with an article, it is then tokenised into sentences.
For every sentence the formulae have to be preprocessed: if there is an acronym in the sentence, it needs to be expanded.
After that, the POS tagger runs and the noun phrases are extracted. The NPs extracted  are sent to the classifier.  The candidates for key phrases and classification codes are laid before to human experts.  Their evaluations are used to improve our machine-learning techniques.
The POS block in the middle of the diagram is really big in terms of complexity.  As said above it, if a new release comes from Stanford, the new block can easily be integrated into our system.

General remark: There are a couple of emerging  machine-learning techniques which have also been used for semantic analysis of documents. They work with Deep Belief Networks (deep Neural Networks) and the outcomes of these experiments are more than promising.

\section{Reviewing services in mathematics}

Reviewing journals have a long tradition in mathematics, and nowadays take the form of electronic databases. Today  zbMATH and MathSciNet~\cite{mathscinet} are the most important bibliographic mathematics databases. They are important tools used by the mathematical community in searching for relevant publications. 
These databases provide the most comprehensive bibliographic information about mathematics enhanced  by a deep content analysis of the publications. 
In the Gutenberg era, all this information was created manually.  
The digital age has changed the situation dramatically. The digitisation of information allows automation like that we are developing to make the production of databases more efficient and uniformly to improve the quality of the database zbMATH.\par 
The math databases have three different layers which are directly geared towards content analysis of a publication:

\begin{itemize}
\item bottom layer: reviews or abstracts
\item second layer: key phrases
\item top level: classification
\end {itemize}

Every layer has its own characteristics, but these layers interact. We will show in the following how we have used reviews or abstracts for key phrase extraction  and classification.

\section{Key phrase extraction in zbMATH}

The relevance of NPs for key pheases identification is also valid for our data. 
The Zentralblatt Math, today the database zbMATH, has reacted to the increasing number of key phrases in the mathematical 
literature and has collected them since the 70s. The field UT {\sl Uncontrolled Terms} was introduced to accentuate single 
terms or phrases) of a publication, e.g.,  {\sl marginal function, quasi-differentiable function, directional differentiability, distance function}. This field lists key phrases created by authors and/or reviewers and/or editors of zbMATH. 
Typically, key phrases of authors will be extended by reviewers and editors within the workflow of zbMATH.
The key phrases presented in zbMATH are searchable (by the specification $ut:$ in the search field) and clickable. 
The key phrases in zbMATH are different in size and quality. 
The current number of all key phrases in the database zbMATH is greater than (not disjunct) 10,200,000 entities. That means, the average number of key phrase  of a publication is not more than 3, which is not sufficient for a description of the content below the MSC level. 

The dominant majority of key phrases in zbMATH are noun phrases including formulae as $C^*$-algebra. Only, a small 
number of the manually created key phrases are single adjectives as key phrases, e.g., {\sl quasiconvex}. No verbs were 
used as key phrases. So, we have focused us to identification of noun phrases (with formulae) as the most important candidates for key phrases until now. 
 
For automatic creation of key phrases, typically only the titles and reviews or abstracts are available. This has the advantage that the number of noun phrases which is the list of candicates is small. Moreover, titles and reviews or abstracts are perfectly suited to detecting and extracting key phrases because they are generally understandable and  summarise the content of a publication in a highly condensed form.\\
But of course, the NLP methods using tokenizing, POS tagging, and chunking, have to be adapted to specific requirements of our data.

\subsection{Problems}

{\bf Relevance:} The NPs extracted are of different values for content analysis. Such phrases as {\sl in the following paper (chapter etc.)} or {\sl an important theorem} are of marginal value for content analysis. Therefore we allocate a weight to each extracted noun phrase. 
A noun phrase is given a {\em very high score} if it is

\begin {itemize}
 
    \item {a named mathematical entity which is defined in a mathematical vocabulary such as Wikipedia, PlanetMath, Encyclopaedia of  
       Mathematics, etc. The number of named mathematical entities in these vocabularies is limited, and not more than 50,000 entities. 
       Typically, such phrases are important in assigning the publication to its mathematical context.}
    \item {identical with a proposed key phrase of the publication.  Most mathematical publications have a (limited) number of key 
       phrases created by the author(s).}
    \item {identical with an existing key phrase in zbMATH. The existing key phrases describe general or special aspects. (The total number of existing key phrases in the database zbMATH is more than 10,200,000, the number of distinct  key phrases is 2,900,000.)}
\end {itemize}
  {\em high score} if it
\begin{itemize} 
 
 \item {contains names of mathematicians: if a noun phrase contains names of mathematicians, it is an indicator that the noun phrase is a name for a special conjecture, theorem, approach or method.} 
 \item {is a acronym: Acronyms are artificial words and have a special spelling. Acronyms are used as abbreviations for longer noun phrases. Acronyms are {\it per se} relevant noun phrases. We compare the extracted candidates for acronyms with our dictionary and resolve them. 
 Generally, the resolution is not unique and depends from the area; some acronyms have up to 20 different meanings.} 
 \item {is or contains specific mathematical formulae: A special mathematical formula in a term, e.g., $H^\infty$-control, is a relevant noun. At least all formulae which are not one-character mathematical notations, are important.}
\end{itemize}
{\em marginal or negative score}
\begin{itemize}
 \item if it provides no additional information about the content. Then, the extracted noun phrase is removed from the candidate list.

 \end{itemize}
{\bf Incomplete Chunking:}
Sometimes, relevant mathematical key  phrases involve a larger number of tokens, e.g., {\sl Browder--Ky Fan and Ky Fan--Glicksberg fixed-point theorems}. Sometimes, the extracted phrases are incomplete. To solve these problems, the rules for chunking  have to be adapted permanently.

\subsection{Processing of NPs:}

In the following, the used methods are listed:
\begin{itemize}

\item Weighting: The weighting of key phrases is done as described above.

\item Redundancy: Very often, some of the extracted NPs are similar. A simple measure for the similarity is the number of different tokens between two phrases. The method used is the LCS (Longest Common Subsequence) algorithm. The NPs are grouped by similarity.

\item Filtering: Groups of similar phrases are replaced by a representative. Selecting a representative is done by using the base vocabulary. Existing key phrases and other resources (e.g., the labels of the MSC classes) are used to select the most suitable phrase.

\item Evaluation by experts: The resulting list of possible key phrases is shown to human experts, e.g.,  editors or reviewers who can remove, change or add phrases.

\end{itemize}

\subsection{Results}

{\bf Number of key phrases and quality:}
In the average, 3 -- 4 key phrases were assigned manually to a publication. The average number of extracted NPs is significantly higher: 10 -- 20 NPs. By the methods described above, the number of candidates is reduced to 7 -- 10 phrases for a publication.
 
Up to now, the evaluation of key phrase extraction by human experts has been started only for particular classes because it means additional expense for human involvement. In the first phase, under 40\% of the phrases were accepted by the experts. The feedback led to a redesign and essential improvements of the methods.  The acceptable proportion of automatically generated key phrases increased to more than 60\% by removing irrelevant phrases. It is planned to integrate the machine-based key phrase extraction in a semiautomatic workflow for zbMATH.

Of course, the quality of the proposed key phrases is dependent on the title and review (abstract).\par

{\bf Controlled vocabulary:}
We applied our tools to the complete zbMATH database. All resulting key phrases and all changes are stored and used for further  enrichment and improvement of key phrases. The set of all positive evaluated key phrases is a first controlled vocabulary of mathematics; the irrelevant noun phrases define the bad list. The first version of the prototype of the controlled vocabulary contains 3,500,000 different phrases. The controlled vocabulary can be structured by topic (MSC classification, see below) and weighted by frequency.\par

\subsection{Further remarks}

{\bf Key phrases and classification:}
The automatically created key phrases were also used for classification as will be described below in detail.  Basing classifiers on the extracted key phrases instead of on reviews has significantly improved the quality of automatic classification.

{\bf  Structuring key phrases:}
Using our method  we get only key phrases which are within a text. For a further enhancement of the key phrases and the controlled vocabulary, we have to know additional relations between the phrases, e.g., synonyms, hypernyms, hyponyms, meronyms. Such ontological relations could be used for structuring and improving the extracted key phrases.

{\bf  Deeper analysis of mathematical formulae:}
Mathematical symbols and formulae form an important part of mathematical publications but they are more important in the full texts of publications than in reviews. \\An analysis of the symbols and formulae found in 
zbMATH has shown that the reviews, or abstracts, contain over 10,000,000 symbols and formulae. Most of them are simple one-character symbols. Nevertheless, the analysis of symbols and formulae and its combination with text 
analysis is of great interest, e.g., the correspondence between a text phrase and a formula seems relevant.  Formulae were integrated in POS tagging and noun phrase extraction as described above. A deeper analysis of mathematical symbols and formulae is planned in cooperation with the MathSearch project.

\section{Classification in zbMATH }

Classification is a well-established concept for organising information and knowledge.
Although it is a well known method, it is not a trivial task. The reasons for difficulties are numerous. Two main reasons are
the classification scheme and the classifying process.

Classes are defined by one or more common properties of the members. Abstracting from individual objects, classification schemes assign the objects to classes.   Classification schemes are not given {\it a priori}, they are intellectually designed and depend on the topic, aims, time, interests and views of the developers of the classification scheme.

The MSC was designed by the American Mathematical Society in the 1970s. The primary goal was to support  subject-oriented access to the increasing number of mathe\-ma\-tics-relevant publications, e.g., zbMATH lists 35,958 journal articles, books and other publications in mathematics and application areas in 1975. For a sufficiently fine-grained access to these thousands of documents annually, a hierarchical three-level deep classification scheme with more than 5,000 classes was developed. In particular, the top level of the MSC has 63 classes.

Typically, the classes of a classification scheme are not pairwise disjoint. Often, an overlapping of classes is part of a concept. This is also valid for the MSC, e.g., Navier-Stokes equations are listed in two main topics: 35-XX Partial differential equations (this is the mathematical point of view) and 76-XX Fluid mechanics (here, the application aspect is dominant). The MSC shows not only hierarchical relations but also different kinds of similarity. Moreover, an object can possess the properties of different classes.
Typically, a mathematical publication cannot be reduced to a unique aspect or property. Publications develop or analyse mathematical models or objects investigated, make quality statements about them, or develop methods or tools to solve problems. This implies that a publication can be a member of more than one class.

A second reason for the difficulties is the classification process. This means that the classification codes (the handles indexing the classes) assigned an object, e.g., the MSC classification codes given a mathematical publication, are subjective (there is a certain range for classification codes). The classifications can be weighted: zbMATH and MathSciNet differentiate between primary and secondary MSC classifications.

What is the true classification of an object?  What is the most important class?  Are the classification codes given complete? Generally, there are often no objective answers to these questions. The uncertainty of the classification scheme and the subjectivity of the human experts work against the objective value of classification.
For the reviewing services in mathematics, authors, editors and reviewers are involved in the classification process. In more detail, the final classification codes in zbMATH are the result of the workflow process. This means, each classification code given by the author(s) is checked by the reviewer and the editor. This workflow reduces the impact of subjective decision.

To the best of our knowledge, there has been no serious analysis of the quality and reliability of assigned classification codes. Checking classification would cover two steps
a.) correctness of a proposed classification code 
b.) completeness of proposed classification codes (often a publication belongs to more than one class, e.g., look for a control system described by ordinary differential equations and its stability is investigated).

But for the development and evaluation of automatic classifiers we are assuming to start with that the classification in zbMATH is correct.

One aim of our work is to provide tools for automatic classifying, especially SVM methods. In detail, we took 63 different SVM classifiers, one for every top-level of the MSC and trained them on key phrases from the corresponding sections of the zbMATH database.

\subsection{Quality}

Quality of classification is usually measured by precision \& recall and the F1-score which is their harmonic mean. Precision is the proportion of all publications of a class which are assigned to a specific class by the automatic classifier. For recall we look at all publications which are not in a class (in the complement of the class: this means the publication is in some other class) but assigned to this class by the automatic classifier. \par

\subsection{Results}

The results of the classification provide a differentiated picture. Roughly speaking; the precision is sufficient (for 26 of the 63 top-level classes the precision is higher than 0.75 and only for 4 classes is it smaller than 0.5), the recall is not.  In other words publications which were classified as elements of a particular MSC class $i$ are mostly correctly classified.  The classifier is precision weighted: for all MSC classes the precision is higher than recall.

In the following, we discuss the results in more detail. A central idea in discussing the quality is the overlapping of classes. Therefore we have built a matrix, indexed by the top-level classes of the MSC, which lists the numbers of publications according to the following definition: 
Let $a_{ii}$ be the number of publications which are classified exclusively with the MSC class $i$
and $a_{ij}$ the number of publications with the primary classification $i$ and secondary classification $j$.

We normalize the elements $a_{ij}$ of the matrix by with $a_{ij}/A_i$ where $A_i$ denotes the number
of all publications with primary MSC code $i$

As a first result, it becomes clear that there is a correlation between the overlapping of classes and the results of the automatic classifier.

\begin{figure}[h!]
\centering
\includegraphics[width=0.4\textwidth]{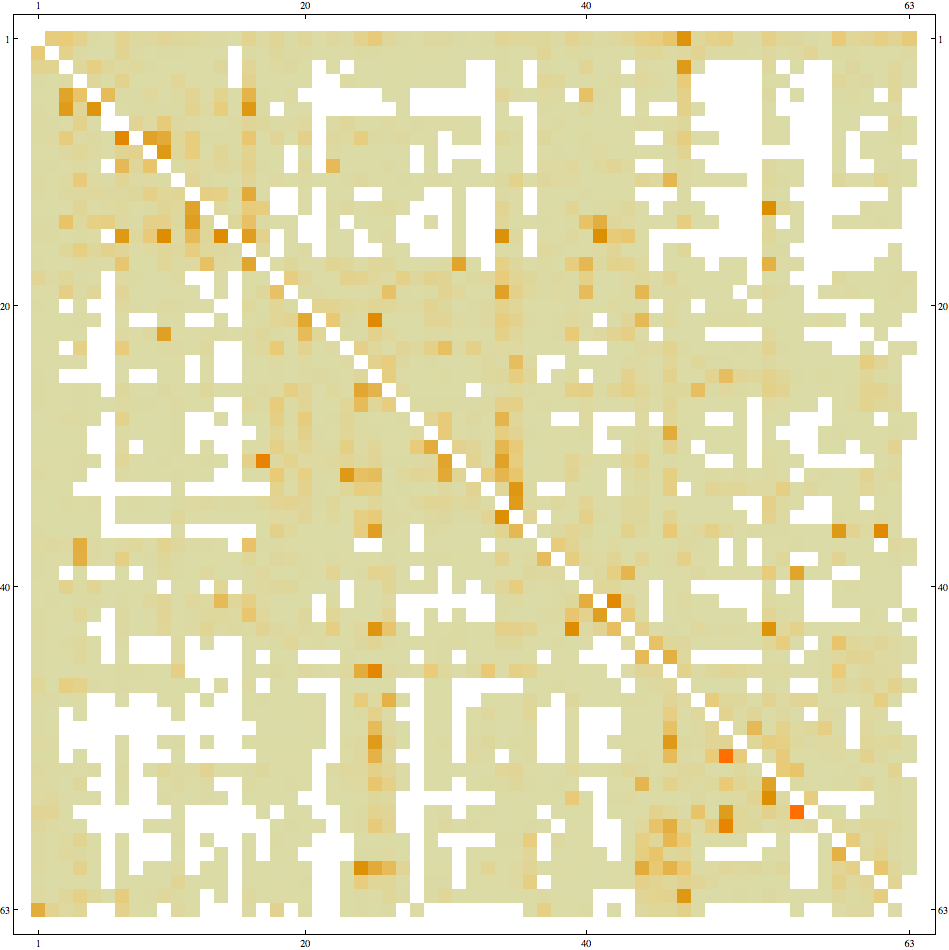}
\caption{The axes are the MSC class numbers, and the entries express the
degrees of overlap $a_{ij}$; white indicates no overlap; the darker
the cell the more the overlap.}
\label{postagbuildblock}
\end{figure}

{\bf Easy classes:} If $a_{ii}/A_i$ is near 1, then both precision and  recall are high. The overlap with other classes is small and the vocabulary differs significantly from the vocabulary of other classes.
Giving examples of such classes are  those publications which have primary classifications from an application areas. This seems to be natural, because each application area has its own specific language and terminology. \par
{\bf Difficult classes:}
There are have different types of overlap. Overlap with other classes can be focused on some MSC classes, e.g., for the class 31-XX {\sl Potential theory} and 43 {\sl Harmonic analysis}, or distributed, e.g., MSC 97-XX {\sl Mathematical education}. In the first case, we propose to further cluster some MSC classes which are similar.

In addition, the total number of all publications with the primary classification $i$ is relevant:  a small number of documents has a negative impact on the classification quality.
It seems that vocabulary and terminology of these classes may not be stable enough. The difficult classes have less than a few thousand documents. \par

\subsection{Remarks}

{\bf Use of classification:}
A high precision means a high reliability that publications of the class $i$ will be also automatically assigned to this class. This is important for preclassification where precision is more important than recall. Until now we have been deploying the classification tool for preclassification of publications.
We propose to improve recall by a second step of classification analysis. The key phrases of each  publication  assigned to the class $i$ by the automatic classifier will be analyzed in more detail.

{\bf  Controlled vocabulary and classification:} Classification is -- in addition to key phrases -- an important piece of metadata in the content analysis of a publication. Each zbMATH item can bear more than one classification code. The database zbMATH does not contain  a relation between key phrases and classification codes. It is a $n:m$ relation. Also the hierarchical structure of the MSC is a problem too. To begin we have applied to top classes and assign a MSC class to a key phrase if the MSC classification (at the top level) is unique. This allows  creating an initial vocabulary for each top MSC class  which has a higher precision of the definition of a MSC class than the existing definition.
Moreover, the structure of the MSC scheme can be analyzed e.g. by the studying the intersection between the controlled vocabularies of different MSC classes.

\section{Conclusion and next steps}

It seems that  the machine-based methods we have developed for key phrase extraction and classification are 
already useful in improving the content analysis of mathematical publications and making the workflow at zbMATH more efficient. We note some positive effects:
\begin{itemize}
\item {Quantity and quality of key phrases is increased by automatic key phrase extraction.} 
\item {The integration of formulae into key phrase extraction lays the foundations for including formulae in content analysis. This could essentially improve content ana\-lysis of mathematical documents.} 
\item {Results of classification can be used to redesign and improve the MSC.} 
\item {The use of standardized methods guarantees a balanced and standardized quality of content analysis in zbMATH.}
\end{itemize}

\end{document}